\documentclass[aps,prd,preprint,superscriptaddress,tightenlines,nofootinbib]{revtex4}
\usepackage{graphicx}
\usepackage{dcolumn}
\usepackage{bm}
\usepackage{epsfig}

\begin{document}

\def\mus{$m_b^{\rm 1S}$}
\def\epm{$e^+e^-$}
\def\ep{$e^+$}
\def\em{$e^-$}
\def\ups{$\Upsilon{\rm 4S}$}
\def\elep{$e^{\pm}$}
\def\mulep{$\mu^{\pm}$}
\newcommand{\nc}{\newcommand}
\newcommand{\rnc}{\renewcommand}
\nc{\bea}{\begin{eqnarray}}
\nc{\eea}{\end{eqnarray}}

\preprint{CLEO CONF 02-10}
\preprint{ICHEP02 ABS932}

\title{Measurement of the lepton energy in the decay
${\bar B} \rightarrow X \ell {\bar \nu}$ and determination of the Heavy Quark
Expansion Parameters}

\thanks{Submitted to the 31$^{\rm st}$ International Conference on High Energy
Physics, July 2002, Amsterdam}


\author{R.~A.~Briere}
\author{G.~P.~Chen}
\author{T.~Ferguson}
\author{G.~Tatishvili}
\author{H.~Vogel}
\affiliation{Carnegie Mellon University, Pittsburgh, Pennsylvania 15213}                    
\author{N.~E.~Adam}
\author{J.~P.~Alexander}
\author{K.~Berkelman}
\author{V.~Boisvert}
\author{D.~G.~Cassel}
\author{P.~S.~Drell}
\author{J.~E.~Duboscq}
\author{K.~M.~Ecklund}
\author{R.~Ehrlich}
\author{L.~Gibbons}
\author{B.~Gittelman}
\author{S.~W.~Gray}
\author{D.~L.~Hartill}
\author{B.~K.~Heltsley}
\author{L.~Hsu}
\author{C.~D.~Jones}
\author{J.~Kandaswamy}
\author{D.~L.~Kreinick}
\author{A.~Magerkurth}
\author{H.~Mahlke-Kr\"uger}
\author{T.~O.~Meyer}
\author{N.~B.~Mistry}
\author{E.~Nordberg}
\author{J.~R.~Patterson}
\author{D.~Peterson}
\author{J.~Pivarski}
\author{D.~Riley}
\author{A.~J.~Sadoff}
\author{H.~Schwarthoff}
\author{M.~R.~Shepherd}
\author{J.~G.~Thayer}
\author{D.~Urner}
\author{G.~Viehhauser}
\author{A.~Warburton}
\author{M.~Weinberger}
\affiliation{Cornell University, Ithaca, New York 14853}                                    
% ...
\collaboration{CLEO Collaboration}
\noaffiliation


\date{July 23, 2002}

\begin{abstract} 

We measured two moments of the lepton momentum spectrum in
${\bar B} \rightarrow X \ell {\bar \nu}$, where $\ell= e$ or $\mu$, for
$p_{\ell}\ge 1.5$ GeV$/c$. From these we derive the Heavy Quark
Expansion (HQE) parameters $\bar{\Lambda}(\overline{\mbox{MS}})=0.39\pm
0.03|_{stat}\pm 0.06|_{sys}\pm 0.12|_{th}$ GeV and $\lambda _1=
-0.25\pm 0.02|_{stat}\pm 0.05|_{sys}\pm 0.14|_{th}$ GeV$^2$, through order
$1/M_B^3$ in the non-perturbative expansion and
$\beta_0\alpha_s^2$ in the perturbative expansion.  The theoretical
expression needed to extract $|V_{cb}|$ from the measured
semileptonic width is evaluated using these HQE parameters.
Combined with the world average of the semileptonic width, we find
$|V_{cb}|=(40.8\pm 0.5|_{\Gamma _{sl}}\pm
0.4|_{(\lambda_1,\bar{\Lambda})_{exp}}\pm 0.9|_{th}) \times
10^{-3}$. Finally, the short range $b$-quark mass \mus\ is evaluated
and found to be $4.82\pm 0.07|_{\exp}\pm 0.11|_{th}$ GeV/$c^2$.

\end{abstract}

\maketitle

\section{Introduction}
Experimental data on inclusive $B$ meson
semileptonic widths have reached great precision \cite{pdg2002}.
In order to use these data to determine the Cabibbo-Kobayashi-Maskawa
(CKM) parameter $|V_{cb}|$, we need a theoretical evaluation 
of the hadronic effects for this process. 
We can make use of theoretical expressions for the differential rate
$d\Gamma_{sl}/dE_\ell$ in terms of $|V_{cb}|$.
The heavy quark
expansion  \cite{manohar-wise,bigi,gremm-kap,secondalph} is a QCD-based 
approach that casts perturbative and non-perturbative
corrections to the partonic width as power series expansions. An
underlying assumption of this approach is quark-hadron duality.

It is important to assess the uncertainty introduced by possible
duality violations quantitatively, in order to achieve a full understanding of
the theoretical errors and be able to ascertain the true uncertainty on $|V_{cb}|$.

The theoretical expression for the inclusive semileptonic width
for ${\bar B} \rightarrow X \ell {\bar \nu}$ ($\ell=\mu$ or $e$)
through ${\cal O}(1/M_B^3)$ in the 
non-perturbative expansion and $\beta_0 (\frac{\alpha_s}{\pi})^2$
in the perturbative one is given by \cite{gremm-kap,vcb1}
\bea \Gamma_{sl} &=&
\frac{{G_F}^2 |{\mbox{V}}_{cb}|^2 {M_B}^{5}}{192 \pi^3} \; 0.3689
\left[ 1-1.54 \frac{\alpha_s}{\pi} -
1.43 \beta_0 \left(\frac{\alpha_s}{\pi}\right)^2 \right. \nonumber \\
 & & - 1.648 \frac{\bar{\Lambda}}{{M_B}} \left(1-0.87 \frac{\alpha_s}{\pi}\right)
- 0.946 \left(\frac{\bar{\Lambda}}{{M_B}}\right)^2
-3.185 \frac{{\lambda}_{1}}{{M_B}^{2}}+0.02\frac{{\lambda}_{2}}{{M_B}^{2}} \nonumber \\
 & &-0.298\left(\frac{\bar{\Lambda}}{{M_B}}\right)^3
-3.28 \frac{{\lambda}_{1} \bar{\Lambda}}{{M_B}^{3}}
 + 10.47 \frac{{\lambda}_{2} \bar{\Lambda}}{{M_B}^{3}}
- 6.153 \frac{{\rho}_{1}}{{M_B}^{3}}+ 7.482 \frac{{\rho}_{2}}{{M_B}^{3}}\nonumber \\
& & \left.- 7.4 \frac{{\tau}_{1}}{{M_B}^{3}} + 1.491 \frac{{\tau}_{2}}{{M_B}^{3}}
 - 10.41 \frac{{\tau}_{3}}{{M_B}^{3}}
 - 7.482 \frac{{\tau}_{4}}{{M_B}^{3}}+ {\cal O}\left(\frac{1}{{M_B}^{4}}\right)
 \right].
\label{vcbf}
\eea
The parameter ${{\lambda}_{1}}$ \cite{manohar-wise,bigi} is
related to the average kinetic energy of the $b$ quark inside the $B$
meson:
\begin{equation}
{{\lambda}_{1}} =\frac{1}{2 M_B} \left< B(v)|{\bar{h_v}} (i D)^2 h_v|B(v)\right>.
\end{equation}
The parameter ${{\lambda}_{2}}$ \cite{manohar-wise,bigi} is the
expectation value of the leading operator that breaks the heavy
quark spin symmetry. It is formally defined as
\begin{equation}
{{\lambda}_{2}} =\frac{-1}{2 M_B} \left<B(v)|{\bar{h_v}} \frac{g}{2} \cdot \sigma^{\mu\nu} 
G_{\mu\nu} h_v|B(v)\right>,
\end{equation}
where $h_v$ is the heavy quark field, and $\left|B(v)\right>$ is the
$B$ meson
state. It is determined from the $B^{\star}-B$ mass difference
to be $0.128\pm 0.010$ GeV$^2$. ${\bar{\Lambda}}$ is related to
the $b$-quark pole mass $m_b$ \cite{manohar-wise,bigi} through the
expression:
\begin{equation}
m_b = {\bar{M}_B}-{\bar{\Lambda}}+\frac{{\lambda}_{1}}{2 m_b},
\end{equation}
where ${\bar{M}_B}$ is the spin-averaged $B^{(\star)}$ mass
(${\bar{M}_B} = 5.313$ GeV$/c^2$). A similar relationship holds between
the $c$-quark mass $m_c$ and the spin-averaged charm meson mass
(${\bar{M}_D} = 1.975$ GeV$/c^2$).

The shape of the lepton energy spectrum in the decays ${\bar B} \rightarrow X \ell {\bar \nu}$
can be used to measure the HQE parameters ${{\lambda}_{1}}$
and ${\bar{\Lambda}}$, through its energy moments, which are also predicted in
the Heavy Quark Expansion. We choose to study truncated moments of the lepton 
spectrum, with a lepton momentum cut of $p_{\ell} \ge 1.5$
GeV$/c$. This choice decreases the 
sensitivity of our measurement to the secondary leptons from the cascade
decays ($b \rightarrow c \rightarrow s/d \ell \nu$). 

The quantities that we focus on, originally suggested by Gremm {\it et
al.} \cite{ligeti},\footnote{Our notation is different than what is
used in Ref.~\protect{\cite{ligeti}},  where R$_0$ is first introduced
as R$_2$.} are
\begin{equation}
{\mbox{R}}_{0} =\frac{\int_{1.7}^{} (d \Gamma_{sl}/dE_l)
dE_l}{\int_{1.5}^{} (d \Gamma_{sl}/dE_l) dE_l},\ {\rm and}
\label{r2}
\end{equation}

\begin{equation}
{\mbox{R}}_{1} =\frac{\int_{1.5}^{} E_l (d \Gamma_{sl}/dE_l)
dE_l}{\int_{1.5}^{} (d {\Gamma_{sl}} /dE_l) dE_l}.
\label{r1}
\end{equation}
The theoretical expressions for these moments R$^{th}$
\cite{gremm-kap,chris} are evaluated by
integrating over the lepton energy in the decay $b\rightarrow c
\ell \bar{\nu}$ for the dominant $\Gamma _c$ component. In
addition, the small contribution coming from charmless
semileptonic decays $b\rightarrow u \ell \bar{\nu}$ is included
in the theoretical expression by adding the integral over $d\Gamma
_{u}/dE_{\ell}$ where the contribution to the moments is scaled by
$|V_{ub}/V_{cb}|^2$ \cite{ligeti,chris}.

We determine these two moments from the measured lepton
spectrum in $\bar{B} \rightarrow X \ell \bar{\nu}$ and insert them in the
theoretical expressions to determine the two parameters $\lambda_1$ and
$\bar{\Lambda}$.  Previous experimental determinations of
$\bar{\Lambda}$ and $\lambda _1$ have been obtained by studying the
$E_\gamma$ spectrum in $b\rightarrow s \gamma$ \cite{bsgamma}
and the first moment
of the mass $M_X$ of the hadronic system recoiling against the
$\ell\nu$ pair in $\bar{B}\rightarrow X \ell \bar{\nu}$
decays~\cite{eht-moments}. We will 
summarize the present knowledge on these parameters and the
implications for $|V_{cb}|$.

In recent years, increasing attention has been focused on ``short-%
range masses,'' preferred by some authors as they are not affected
by renormalon ambiguities \cite{ikaros-kolya}.  In particular,
theoretical determinations of the so-called $b$-quark mass in the $1S$
renormalization scheme, \mus, based on
extractions from $\Upsilon$(1S) resonance data are quoted with
impressive theoretical accuracy \cite{h1}. Using the formalism
developed by C.~Bauer and M.~Trott \cite{chris}, we have used
semileptonic energy moments to extract \mus .

\section{Experimental Method}
The data sample used in this study was collected with the CLEO II
detector \cite{cleoii} at the Cornell \epm\ collider (CESR). It
consists of an integrated luminosity of 3.14 fb$^{-1}$ at the
$\Upsilon$(4S) energy, corresponding to a sample of $3.3\times 10^{6}$
$B\bar B$ events. The continuum background is studied with a sample of
$1.61$ fb$^{-1}$ collected at an energy about $60$ MeV below the
resonance.

We measure the momentum spectrum of electrons and muons with a minimum
momentum requirement of 1.5 GeV$/c$.  This momentum cut
serves two purposes. First, it ensures that we can take
advantage of good particle identification criteria for both lepton
species.  Having both lepton species allows us to check crucial
systematic effects by examining the $\mu/e$ ratio.
In addition, in this range the inclusive spectra are
dominated by the direct $b\rightarrow c\ell \nu$ semileptonic
decay, rather than by secondary leptons produced in the decay
chain $b\rightarrow c \rightarrow s \ell \nu$. 

Electrons are identified with a likelihood method that includes several
discriminating variables, most importantly the ratio $E/p$ of the
energy deposited in the electromagnetic calorimeter to the
reconstructed momentum, and the specific
ionization in the central drift chamber.  Muon candidates are required
to penetrate at least five nuclear interaction lengths of absorber material.
 We use the central part of the detector
($|\cos{\theta}| \le 0.71$ for electrons and $|\cos{\theta}| \le 0.61$
for muons). 

The overall efficiency is the product of three factors: the
reconstruction efficiency, including event selection
criteria and acceptance corrections; the tracking efficiency; and
the $\mu$ or $e$ identification efficiency. The first two
factors are close to 1 and estimated from Monte Carlo simulations and
checked with data,
whereas the lepton identification efficiencies are studied with
data: radiative $\mu$-pair events for the $\mu$ efficiency and
radiative Bhabha electron tracks embedded in hadronic events for the
$e$ efficiency. 
The $e$ identification efficiency is rather constant in our
momentum range and equal to $93.8 \pm 2.6 \%$.  The $\mu$ momentum 
threshold is near our low momentum cut, but the efficiency rises to 
a plateau of about 95\% above 2.0 GeV$/c$.

\begin{figure}[htbp]
\center{
{\epsfig{figure=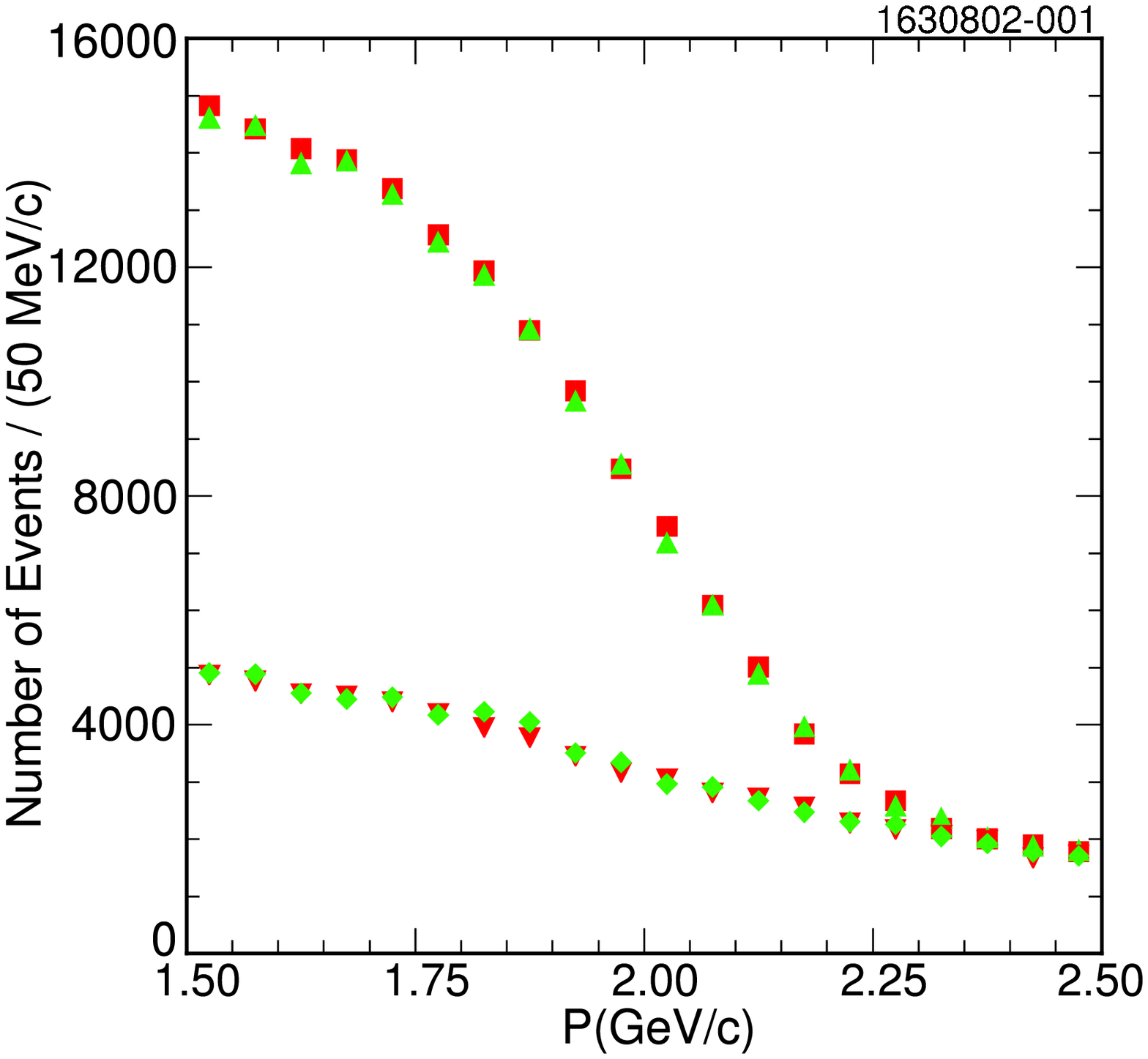,width=4in,height=3.5in}}
{\epsfig{figure=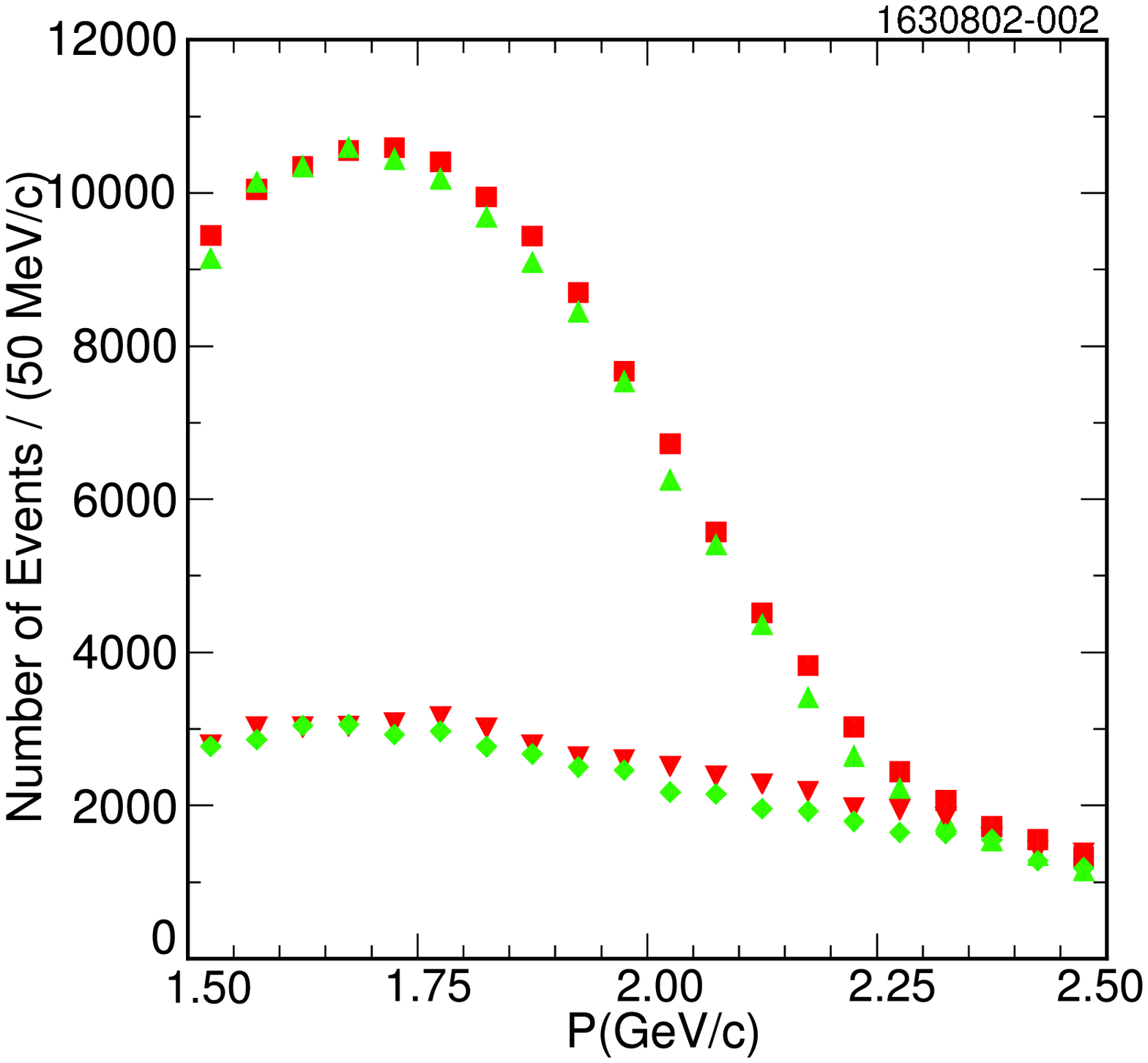,width=4in,height=3.5in}}
}
\caption{\small {Raw electron (top) and muon (bottom) spectra from
$\Upsilon(4S)$ (\hbox{\vrule height8pt width8pt depth2pt} and $\triangle$ ) and scaled continuum ($\diamond$
and $\nabla$) data.  Positive (red) and negative (green) leptons are
shown separately.}}
\label{totelspec}
\end{figure}

Figure \ref{totelspec} shows the raw yields
for electrons (top) and muons (bottom) from the $\Upsilon$(4S) and
the scaled continuum background. The scaling factor for the
continuum sample is determined by the ratio of integrated
luminosities and continuum cross sections and is $1.930\pm 0.013$.
This scale factor has been determined independently using high
statistics samples of tracks with momenta higher than the
kinematic limit for $B$-meson decay products. This control
sample has been used to determine the systematic errors in the
measurement. 

The raw yields include hadrons misidentified as
leptons (fakes). This contribution is determined from data as follows.
Fake rates are determined from tagged samples:
$\pi$'s from $K^0_S\rightarrow \pi^+\pi^-$, 
$K$'s from $D^{*+} \rightarrow D^0 \pi^+, D^0 \rightarrow K^-\pi^+$, 
and $p$'s from $\Lambda \rightarrow p \pi^-$. 
Using the measured fake rates and the observed spectrum of 
hadronic tracks, we compute a fake contribution, which is quite small
for both electrons and muons.

Several sources of real leptons must be corrected for. Leptons from
$J/\psi$ decays are vetoed by combining a candidate with another
lepton of the same type and opposite sign and removing it if their
invariant mass is 
within $3\sigma$ of the known $J/\psi$ mass. Corrections are applied
for veto inefficiency and for signal leakage into the veto window.
Similar procedures are applied to electrons and positrons coming
from $\pi^0$ Dalitz decays and from $\gamma$ conversions. 

\begin{figure}[htbp]
\center{
{\epsfig{figure=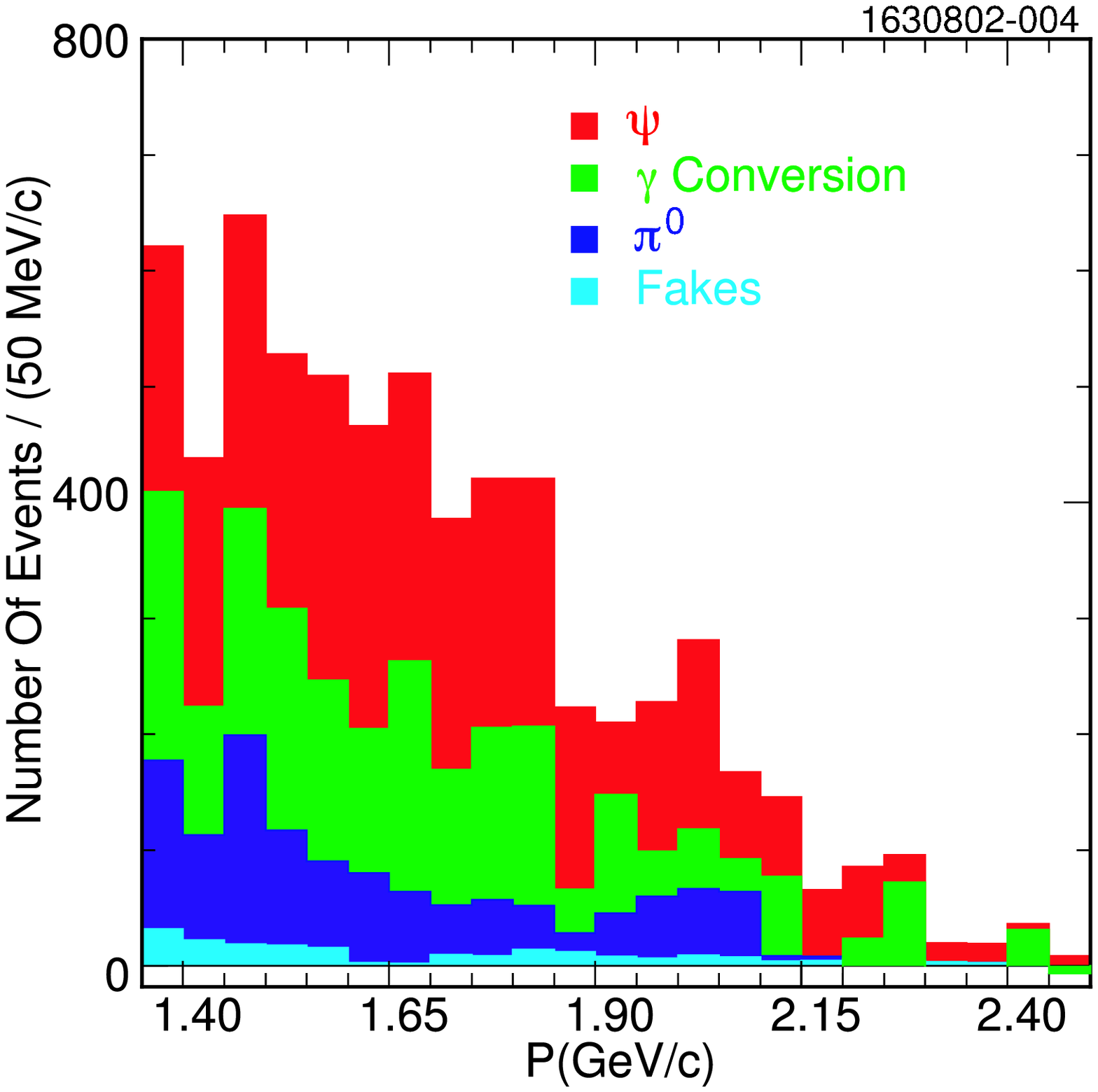,width=4in,height=3.5in}}
{\epsfig{figure=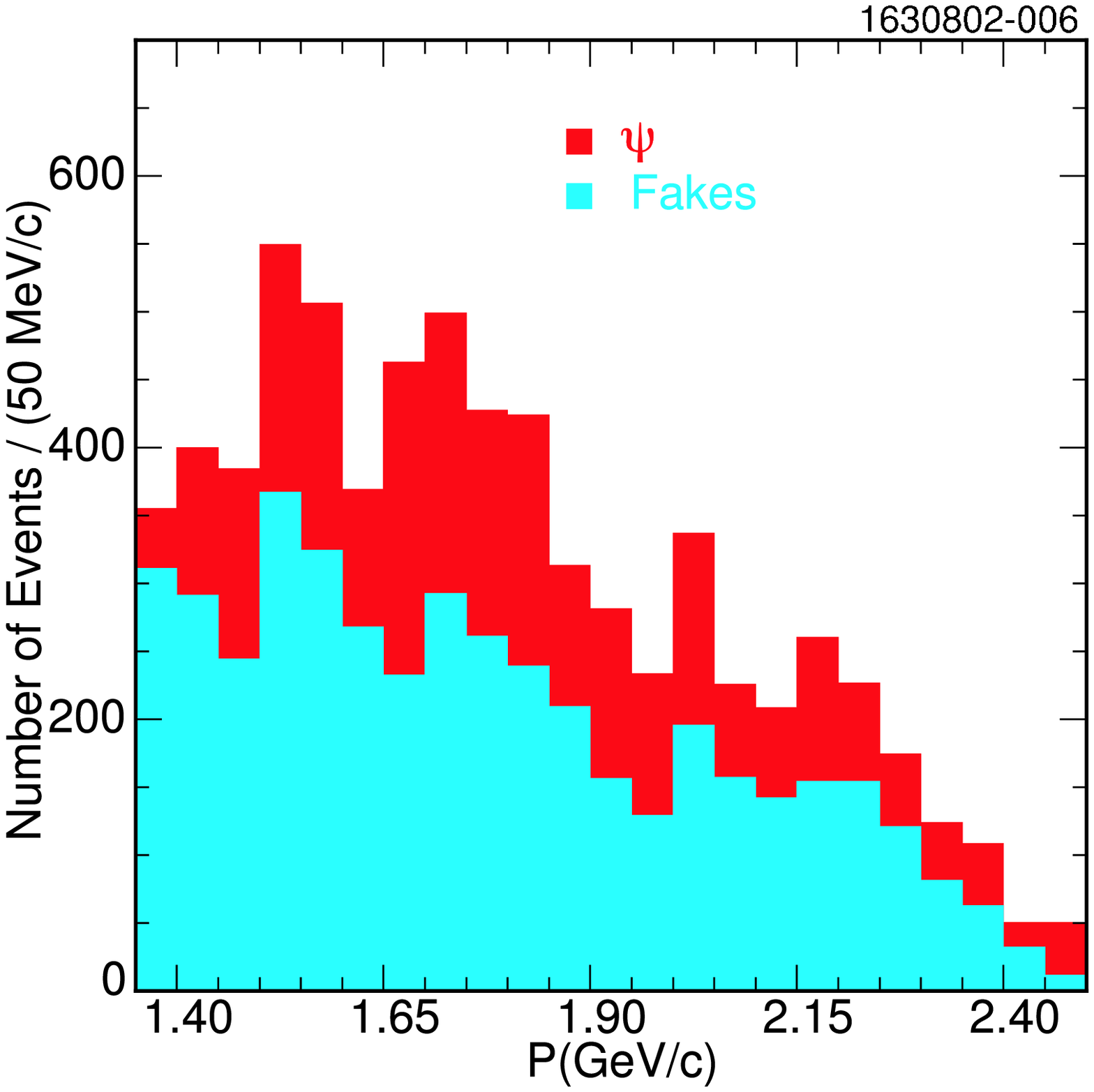,width=4in,height=3.5in}}
}
\caption{\small {
Contributions to the electron (top) and muon (bottom) spectra from
background processes that were estimated with data.
}}
\label{elbkgcntri}
\end{figure}

\begin{figure}[htbp]
\center{
{\epsfig{figure=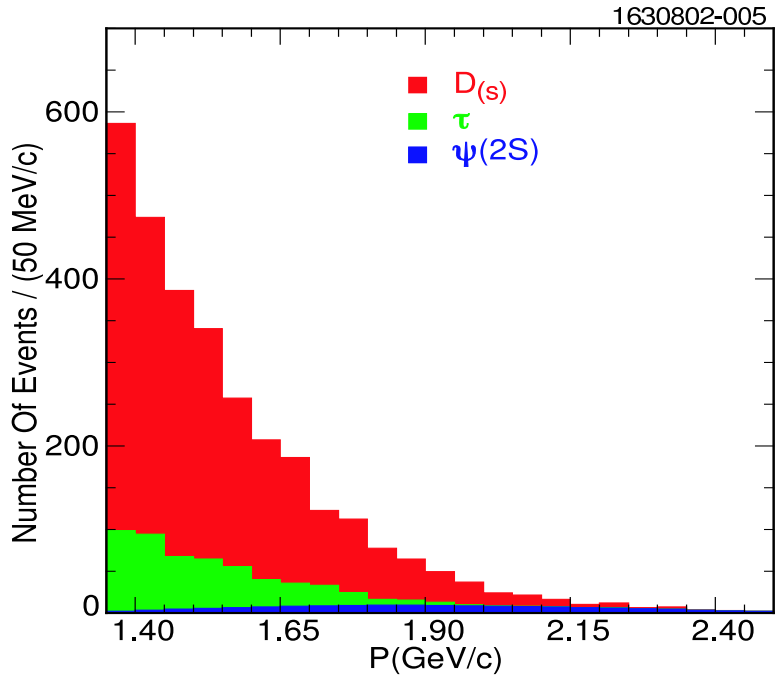,width=4in,height=3.5in}}
}
\caption{\small {
Contributions to the electron spectrum from background sources that
were modeled with Monte Carlo.
}}
\label{elsec}
\end{figure}

Finally, we subtract leptons coming from $\psi^\prime$ decays or the
secondary decays  $b\rightarrow c\rightarrow s \ell \nu$ and
$B\rightarrow \tau \rightarrow \ell \nu \bar{\nu}$  
using Monte Carlo studies. Figures \ref{elbkgcntri} and \ref{elsec}
show the individual estimated background contributions to our sample.
Note that all of the backgrounds are small compared to the signal.

Our goal is a precise determination of the shape of the lepton
momentum spectrum, so corrections for the distortion introduced by
electroweak radiative effects are important. We use the prescription
developed by D.~Atwood and W.~Marciano \cite{atw-marc}. This procedure
incorporates leading-log and short-distance loop corrections, and sums
soft-virtual and real-photon corrections to all orders. It does not
incorporate hard-photon bremsstrahlung, which mainly modifies the low
energy portion of the electron spectrum, and is not used in our
analysis. An independent method of studying QED radiative corrections
in semileptonic decays, based on the simulation package PHOTOS
\cite{photos}, has been used to obtain an independent assessment of the
corrections.  The difference between the two methods is used to obtain
the systematic error of this correction.

\begin{figure}[htbp]
\center{
{\epsfig{figure=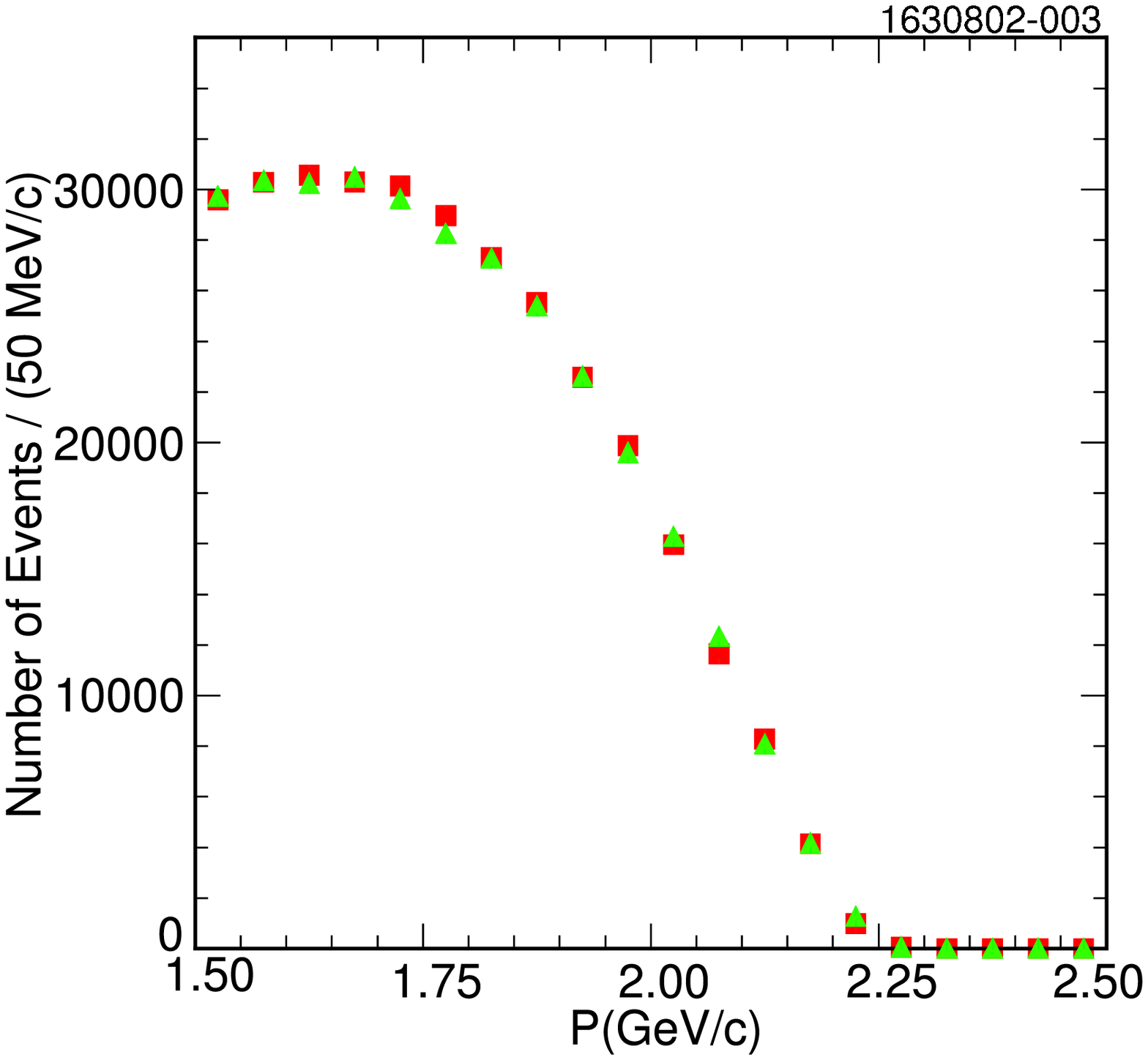,width=4.in,height=3.5in}}
}
\caption{\small {Corrected electron (green $\triangle$) and muon 
(red \hbox{\vrule height8pt width8pt depth2pt}) momentum spectra in the
$B$-meson rest frame.}} 
\label{bothspcatw}
\end{figure}

Finally, we use a Monte Carlo sample of signal events to derive an
unfolding matrix to boost the spectra into the $B$-meson rest frame
($B$ mesons typically have a momentum of $\approx$ 300 MeV$/c$ in the
laboratory frame). Figure \ref{bothspcatw} shows the resulting
electron and muon spectra.

Our first goal is to extract the truncated moments R$_0$ and R$_1$
defined in Eqs.~\ref{r2} and \ref{r1}. Using the measured
spectra, we evaluate 
the relevant integrals and obtain the results shown in
Table~\ref{rmom}, where the first error is statistical and the second
is systematic in each quoted number.
\begin{table}[h]
\begin{center}
\begin{tabular}{|l|c|c|}
\hline

         &   R$^{exp}_0$                 &   R$^{exp}_1$                 \\ \hline
\elep\   &$0.6184 \pm 0.0016 \pm 0.0017$ &$1.7817 \pm 0.0008 \pm 0.0010$ \\ \hline
\mulep\  &$0.6189 \pm 0.0023 \pm 0.0020$ &$1.7802 \pm 0.0011 \pm 0.0011$ \\ \hline \hline
Combined &$0.6187 \pm 0.0014 \pm 0.0016$ &$1.7810 \pm 0.0007 \pm 0.0009$ \\ \hline
\end{tabular}
\end{center}
\caption{\label{rmom} Measured truncated lepton moments for \elep\ and
\mulep, and for the sum.}
\end{table}

The dominant uncertainty for both lepton species is related to
particle identification efficiency. As the moments are ratios of
measured quantities, the effects of several uncertainties, which
are nearly independent of the of the energy, cancel out. The
overall systematic uncertainties are 0.28\% for R$^{exp}_0$ and 0.06\% for
R$^{exp}_1$ for the \elep\ sample, and 0.32\% and 0.06\% for the
\mulep\ sample. These are comparable to the 
statistical uncertainties.  Since the two moments are extracted from
the same spectra, we must use the covariance matrix V$_{\alpha
\beta}$ to extract the HQE parameters.  Table \ref{correlation} shows
this covariance matrix for electrons and muons.

\begin{table}[h]
\begin{center}
\begin{tabular}{|l|c|} \hline
                & V$_{\alpha \beta}$ \\ \hline
Electrons &
$\left( \begin{array}{cc}
2.4 \cdot 10^{-06} & 1.1 \cdot 10^{-06} \\
1.1 \cdot 10^{-06} & 6.7 \cdot 10^{-07}
\end{array} \right)$

\\ \hline

Muons     &

$\left( \begin{array}{cc}
5.2 \cdot 10^{-06} & 2.2 \cdot 10^{-06} \\
2.2 \cdot 10^{-06} & 1.3 \cdot 10^{-06}
\end{array} \right)$
\\ \hline \hline

Combined   & 

$\left( \begin{array}{cc}
1.9 \cdot 10^{-06} & 8.4 \cdot 10^{-07} \\
8.4 \cdot 10^{-07} & 5.0 \cdot 10^{-07}
\end{array} \right)$
\\ \hline

\end{tabular}
\caption{Covariance matrices for the statistical errors on R$^{exp}_0$
and R$^{exp}_1$ moments.} \label{correlation}
\end{center}
\end{table}

\section{Determination of the HQE parameters $\bar{\Lambda}$ and $\lambda _1$}

We can determine $\bar{\Lambda}$ and $\lambda_1$
using the published expressions for the moments R$_0$ and R$_1$ in
terms of the HQE parameters \cite{ligeti,chris}. We use the terms
describing the integral over the $b\rightarrow q \ell \bar{\nu}$ processes.
These theoretical expressions include terms accounting  for
electroweak radiative effects and unfolding from the laboratory to
the rest frame.  We do not include these terms because our data
are corrected for these effects. The theoretical expressions
include terms through order $1/M_B^3$ in the non-perturbative
expansion \cite{manohar-wise,bigi,gremm-kap}. These introduce $6$
additional form factors, denoted as $\rho_1$, $\rho _2$, $\tau _1$,
$\tau _2$, $\tau _3$, and $\tau _4$. From dimensional arguments,
they are expected to be of the order $\Lambda _{QCD}^3$ and thus
they are generally assumed to be $\leq (0.5)^{3}$ GeV$^3$. In
addition, $\rho _1$ is expected to be positive from the vacuum-saturation
approximation \cite{ikaros-rho1}. Furthermore, as Gremm and
Kapustin have noted \cite{gremm-kap}, the $B^*$-$B$ and $D^*-D$ mass
splittings impose the constraint
\begin{equation}
\rho _2-\tau _2 -\tau _4= \frac{\kappa(m_c)M_B^2 \Delta M_B (M_D +
\bar{\Lambda})- M_D^2\Delta M_D (m_B+\bar{\Lambda})}
{M_B+\bar{\Lambda} - \kappa(m_c) \cdot (M_D+\bar{\Lambda})},
\end{equation}
where $\kappa(m_c)\equiv [\alpha _s(m_c)/\alpha
_s(m_b)]^{(3/\beta_0)}$ and $\Delta M_{B,D}$ represents the
vector-pseudoscalar meson splitting in the charm and beauty
sectors.

\begin{figure}[ht]
\center{\epsfig{figure=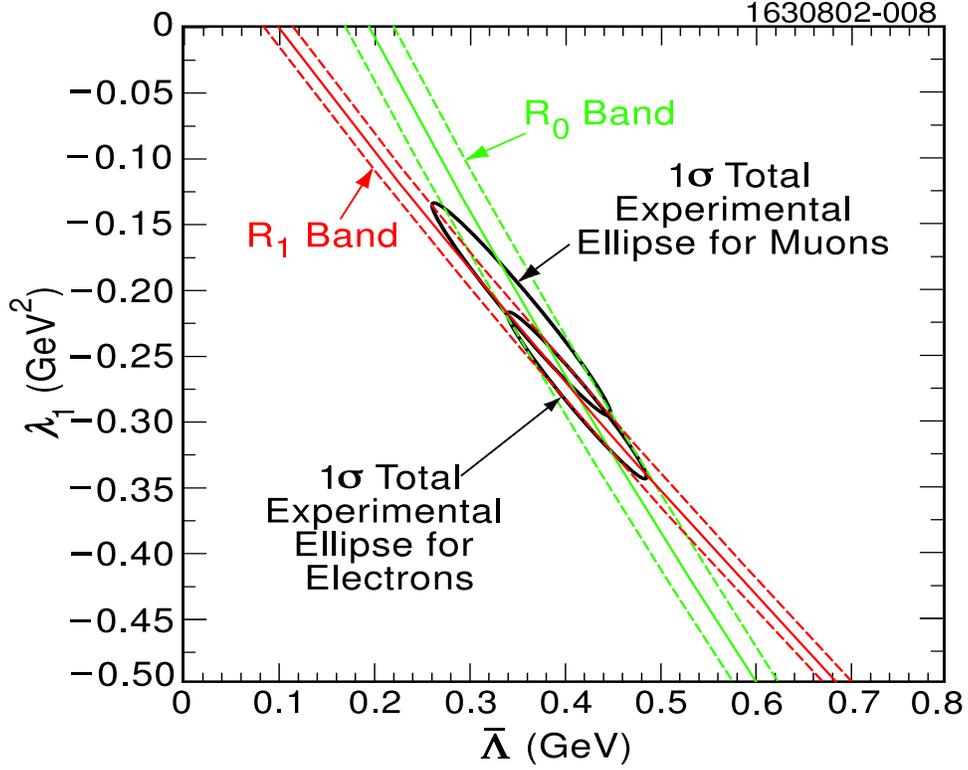,width=5.0in,height=4in}}
\caption{
Constraints on the HQE parameters $\lambda_1$ and $\bar\Lambda$ from
our measurements of the moments R$_0$ and R$_1$.  The contours
represent $\Delta\chi^2=1$ for the combined statistical and systematic
errors on the measured values.  The parameters $\lambda_1$ and $\bar
\Lambda$ are computed in the $\overline{MS}$ scheme to order $1/M^3_B$
and $\beta_0 \alpha_s^2$.
} 
\label{elpsedata}
\end{figure}

The values of the HQE parameters and their 
experimental uncertainties are obtained by calculating the $\chi^2$
from the measured moments R$^{exp}_0$ and R$^{exp}_1$ and the
covariance matrix $\mbox{V}_{\alpha \beta}$:
\begin{equation}
\chi^2= \sum_{\alpha=0}^{\alpha=1} \sum_{\beta=0}^{\beta=1}
(\mbox{R}^{exp}_{\alpha}-\mbox{R}^{th}_{\alpha}) \; \;
\mbox{V}^{-1}_{\alpha \beta}  \; \;
(\mbox{R}^{exp}_{\beta}-\mbox{R}^{th}_{\beta}).
\end{equation}
In Fig. \ref{elpsedata} we show the $\Delta\chi^2=1$ contours for
electrons and muons corresponding to the quoted experimental
uncertainties.

\begin{figure}[ht]
\center{\epsfig{figure=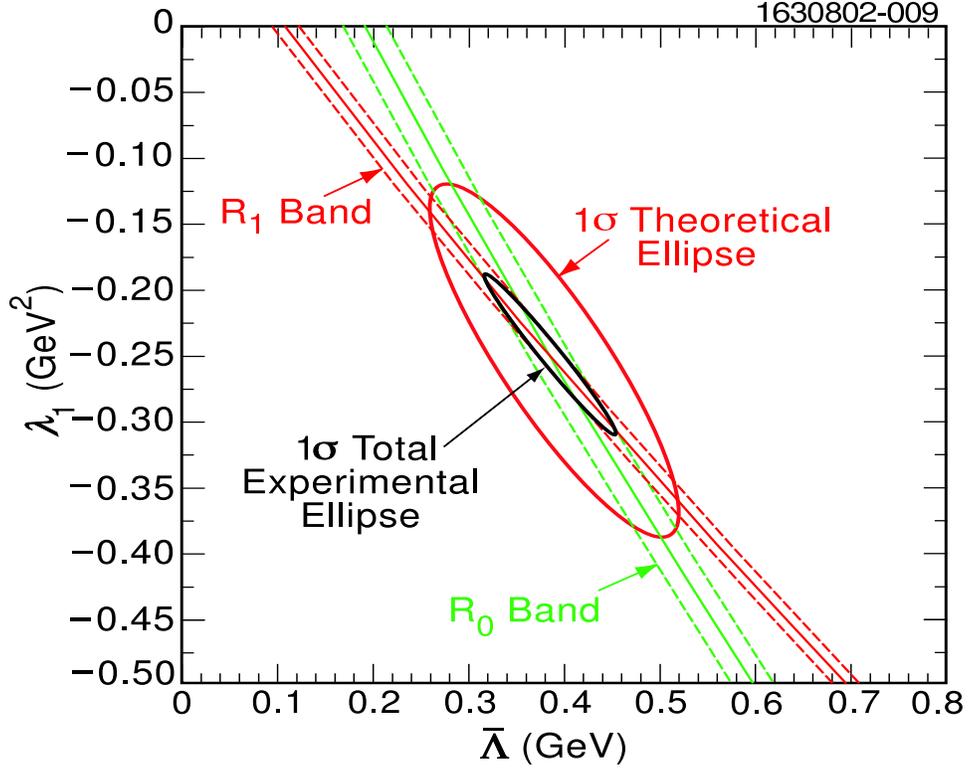,width=5.in,height=4.in}}
\caption{The constraints from our combined electron and muon R$_0$ and
R$_1$ moments, with $\Delta\chi^2=1$ contours for total experimental
and theoretical uncertainties. The parameters $\lambda_1$ and
$\bar \Lambda$ are computed in the $\overline{MS}$ scheme to order
$1/M^3_B$ and $\beta_0 \alpha_s^2$.}
\label{therrepse}
\end{figure}

The theoretical uncertainties on the HQE parameters are determined
by varying, with flat distributions, the input parameters within
their respective errors: $|\frac{V_{ub}}{V_{cb}}|= 0.09 \pm
0.02$, $\alpha_s = 0.22 \pm 0.027$, $\lambda_{2} = 0.128 \pm
0.010$ GeV$^2$, $\rho_1 = \frac{1}{2}(0.5)^3 \pm \frac{1}{2}(0.5)^3$
GeV$^3$, $\rho_2 = 0 \pm (0.5)^3$ GeV$^3$, $\tau_i = 0.0 \pm
(0.5)^3$ GeV$^3$.  The contour that contains 68\% of the
events is shown in Fig. \ref{therrepse}. This procedure for evaluating
the theoretical uncertainty from the unknown expansion parameters that 
enter at order $1/M_B^3$ is similar to that used by Gremm and Kapustin
\cite{gremm-kap}  and Bauer and Trott \cite{chris}, but different from
the procedure used in our analysis of hadronic mass moments
\cite{eht-moments}. The dominant theoretical 
uncertainty is related to the $1/M_B^3$ terms in the
non-perturbative expansion discussed before.

The measured ${{\lambda}_{1}}$ and ${\bar{\Lambda}}$ are given in
Table \ref{statsyst}. 

\begin{table}[h]
\begin{center}
\begin{tabular}{|l|c|c|c|c|} \hline

                & ${\lambda}_{1}$(GeV$^2$)  & $\bar{\Lambda}$(GeV)  \\ \hline
Electrons&$-0.28\pm0.03|_{stat}\pm0.06|_{syst}\pm0.14|_{th}$
&$0.41\pm 0.04|_{stat}\pm 0.06|_{syst}\pm0.12|_{th}$\\ \hline
Muons    &$-0.22\pm0.04|_{stat}\pm0.07|_{syst}\pm0.14|_{th}$
&$0.36\pm 0.06|_{stat}\pm 0.08|_{syst}\pm0.12|_{th}$\\ \hline \hline

Combined &$-0.25\pm0.02|_{stat}\pm0.05|_{syst}\pm0.14|_{th}$
&$0.39\pm 0.03|_{stat}\pm 0.06|_{syst}\pm0.12|_{th}$\\ \hline

\end{tabular}
\caption{Measured ${\lambda}_{1}$ and $\bar{\Lambda}$ values,
including statistical, systematic, and theoretical errors.}
\label{statsyst}
\end{center}
\end{table}

\begin{figure}[ht]
\center{\epsfig{figure=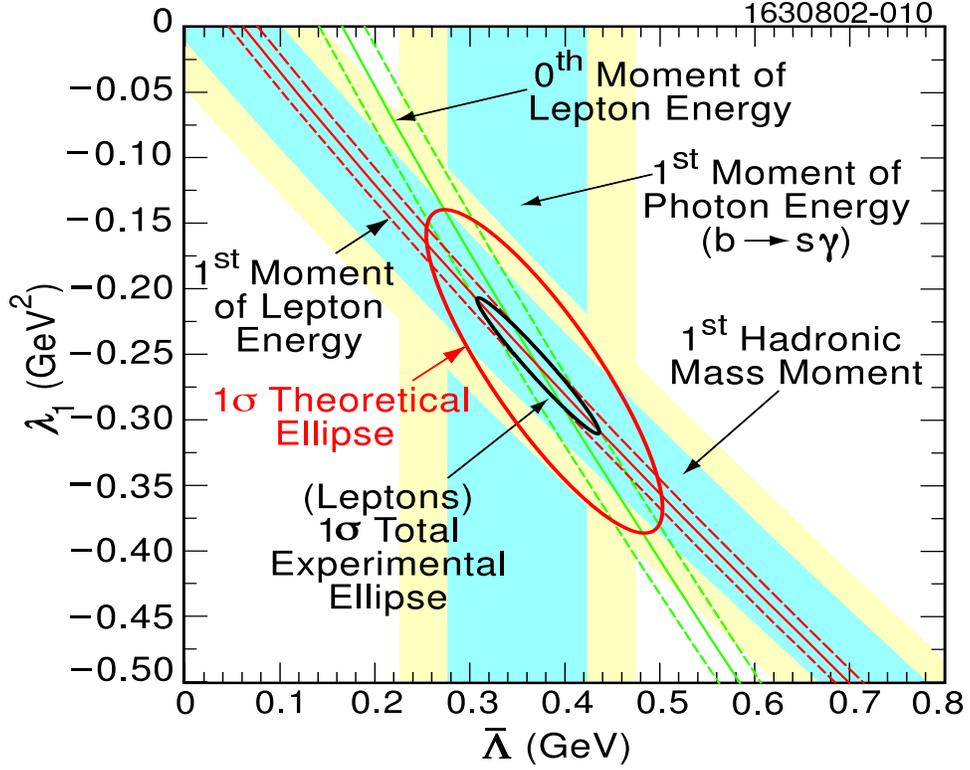,width=5.in,height=4.in}}
\caption{Constraints from the ${\bar B}\to X_c\ell{\bar\nu}$ hadronic
mass moment and $b\rightarrow s \gamma$ $E_\gamma$ moment
\cite{eht-moments} compared with the combined electron and muon R$_0$
and R$_1$ constraints.  
The parameters $\lambda_1$ and $\bar \Lambda$ are
computed in the $\overline{MS}$ scheme to order $1/M^3_B$ and $\beta_0
\alpha_s^2$.} 
\label{hadbsglep}
\end{figure}

A previous CLEO measurement used the first moment of the hadronic
recoil mass \cite{eht-moments} 
and the first photon energy moment from the $b \rightarrow s \gamma$
process \cite{bsgamma}. Figure \ref{hadbsglep} shows a comparison of our results
with the previously published ones. We overlay the ellipse from the electron spectrum
measurements extracted using $|\frac{V_{ub}}{V_{cb}}|= 0.07$ for consistency with their
assumptions. The agreement is good, although the theoretical uncertainties do not warrant a very
precise comparison.

Using the expression of the full semileptonic decay width given in
Eq.~\ref{vcbf}, 
we can extract $|{\mbox{V}}_{cb}|$. We use $\Gamma_{sl}^{exp}=(0.43
\pm 0.01){\times} 10^{-10}$MeV \cite{artuso}.  Assuming the validity
of quark-hadron duality, 

\begin{equation}
|{\mbox{V}}_{cb}|=(40.8\pm 0.5|_{\Gamma_{sl}}\pm0.4|_{\lambda_1, \bar{\Lambda}}\pm 0.9|_{th})\times 10^{-3},
\end{equation}
where the first uncertainty is from the experimental
value of the semileptonic width, the second uncertainty is from the
HQE parameters (${\lambda}_{1}$ and $\bar{\Lambda}$) and the third 
uncertainty is the theoretical uncertainty obtained as described above.

Finally, we use the formalism of Ref. \cite{chris} to extract
the short range mass of the $b$-quark \mus , defined as
$m_b^{1S}\equiv {\bar M}_B- {\bar{\Lambda}}^{1S}$(GeV). Table \ref{lamb1s}
summarizes the measurement of these parameters for electrons, muons
and the sum. The theoretical uncertainty is extracted using the 
method described above.
Our result is in good agreement with a previous estimate of
$m_b^{1S}$ \cite{h1} derived from $\Upsilon (1S)$ data,
\mus$=4.69 \pm 0.03$ GeV$/c^2$.

\begin{table}[h]
\begin{center}
\begin{tabular}{|l|c|c|c|} \hline
           &${\bar{\Lambda}}^{1 S}$(GeV)  
                       &$m_b^{1S}$(GeV$/c^2$) \\ \hline
Electrons&$0.52\pm0.04|_{stat}\pm0.06|_{syst}\pm0.11|_{th}$
                       & $ 4.79 \pm 0.07|_{exp}\pm 0.11|_{th}$\\ \hline
Muons    &$0.46\pm0.05|_{stat}\pm0.08|_{syst}\pm0.11|_{th}$
                       & $ 4.85 \pm 0.09|_{exp}\pm 0.11|_{th}$\\ \hline \hline
Combined &$0.49\pm0.03|_{stat}\pm0.06|_{syst}\pm0.11|_{th}$
                       & $ 4.82 \pm 0.07|_{exp}\pm 0.11|_{th}$\\ \hline
\end{tabular}
\caption{The measured ${\bar{\Lambda}}^{1S}$ and $m_b^{1S}$. The
quoted errors reflect statistical, systematic and theoretical uncertainties,
respectively.} \label{lamb1s}
\end{center}
\end{table}

\section{Conclusion}

We have measured the moments R$_0$ and R$_1$ of the
lepton energy in ${\bar B} \rightarrow X \ell {\bar \nu}$ decays and obtained the
HQE parameters $\bar{\Lambda}= 0.39 \pm 0.03|_{stat} \pm
0.06|_{syst} \pm 0.12|_{th}$ GeV and ${\lambda}_{1}= -0.25 \pm
0.02|_{stat} \pm 0.05|_{syst}\pm 0.14|_{th}$ GeV$^2$. These
results imply that $m_b^{pole} = 4.90 \pm 0.08|_{exp}\pm
0.13|_{th}$ GeV/$c^2$. The short range mass $m_b^{1S}$ is found to be
$4.82 \pm 0.07|_{exp}\pm 0.11|_{th} $ GeV/$c^2$. Finally, we obtain
$|{\mbox{V}}_{cb}|=(40.8\pm0.5|_{\Gamma_{sl}}\pm0.4|_{\lambda_1,
\bar{\Lambda}}\pm0.9|_{th})\times 10^{-3}$.

These results are consistent with previous measurements, and thus we
have no evidence for quark-hadron duality violations. A quantitative
assessment of these unknown errors will be possible only when the
$1/M_B^3$ uncertainties are known more precisely. 

\section{Acknowledgements}
We gratefully acknowledge the effort of the CESR staff in providing us with
excellent luminosity and running conditions.
M.~Selen thanks the PFF program of the NSF and the Research Corporation, 
and A.~H.~Mahmood thanks the Texas Advanced Research Program.
This work was supported by the National Science Foundation and the
U.~S.~Department of Energy. 

We would like also to thank Z.~Ligeti, C.~Bauer, and N.~G.~Uraltsev for 
helpful discussions.

\newpage

\begin{figure}[ht]
\center{\epsfig{figure=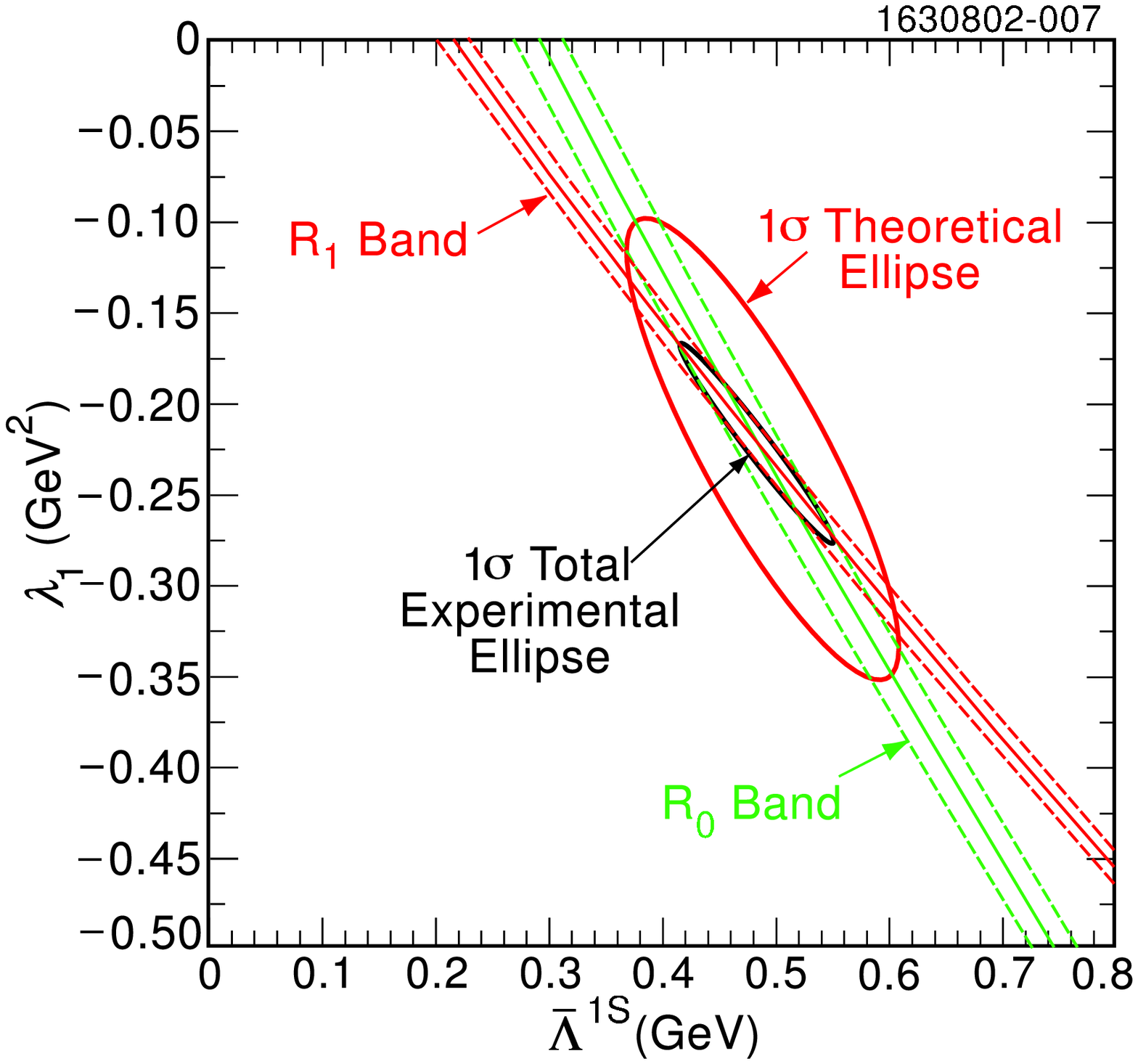,width=5.in,height=4.in}}
\caption{The combined electron and muon R$_0$ and R$_1$ constraints, with
$\Delta\chi^2=1$ contours for total experimental and theoretical
uncertainties in the plane of $\lambda_1$ and $\bar\Lambda^{1S}$,
according to the scheme of Ref.~\cite{chris} to order $1/M_B^3$.}
\label{thr01pse}
\end{figure}


\newpage

\begin{thebibliography}{99}

\bibitem{pdg2002} K.~Hagiwara {\it et al.} (Particle Data Group), Phys. Rev. D {\bf 66}, 010001 (2002).

\bibitem{manohar-wise}
A.~V.~Manohar and M.~B.~Wise, Phys. Rev. D {\bf{49}}, 1310 (1994). 

\bibitem{bigi} 
I.~Bigi, M.~A.~Shifman, N.~G.~Uraltsev, and A.~Vainstein, Phys. Rev. Lett.
{\bf 71}, 496 (1993).

\bibitem{gremm-kap}
M.~Gremm and A.~Kapustin, Phys. Rev. D {\bf{55}}, 6924, (1997).

\bibitem{secondalph}
M.~Gremm and I.~Stewart,  Phys. Rev. D {\bf{55}}, 1226, (1997).

\bibitem{vcb1}
I.~Bigi, N.~G.~Uraltsev, and A.~Vainstein, Phys. Lett. B {\bf 293},
430 (1992); {\it Erratum}, {\bf 297}, 477 (1993); 
M.~Jezabeck and J.~H.~K\"{u}hn, Nucl. Phys. {\bf B314}, 1 (1989);
M.~Luke, M.~J.~Savage, and M.~B.~Wise, Phys. Lett. B {\bf 345}, 301 (1995).

\bibitem{ligeti} 
M.~Gremm, A.~Kapustin, Z.~Ligeti, and M.~Wise,
Phys. Rev. Lett. {\bf{77}}, 20, (1996).

\bibitem{chris} C.~Bauer and M.~Trott, hep-ph/0205039.

\bibitem{bsgamma}
CLEO Collaboration, S.~Chen {\it et al.}, 
Phys. Rev. Lett. {\bf 87}, 251807 (2001).

\bibitem{eht-moments}
CLEO Collaboration, D.~Cronin-Hennessy {\it et al.},
Phys. Rev. Lett. {\bf 87}, 251808 (2001).

\bibitem{ikaros-kolya}
I.~I.~Bigi, M.~Shifman, N.~Uraltsev, and A.~I.~Vainshtein, Phys. Rev. D
{\bf 50}, 2234 (1994).

\bibitem{h1}
A.~H.~Hoang, Phys. Rev. D {\bf 61}, 034005 (2000) [hep-ph/9905550];
Nucl. Phys. B, Proc. Suppl. {\bf 86}, 512 (2000) [hep-ph/9909356];
Phys. Lett. B {\bf 483}, 94 (2000) [hep-ph/9911461];
CERN-TH/2000-227 [hep-ph/0008102];
CERH-TH/2001-48 [hep-ph/0102292].

\bibitem{cleoii}
CLEO Collaboration, Y.~Kubota {\it et al.}, Nucl. Instrum. Meth. A
{\bf 320}, 66 (1992).

\bibitem{atw-marc}
D.~Atwood and W.~Marciano, Phys. Rev. D {\bf{41}}, 1736 (1990).

\bibitem{photos}
E.~Barberio and Z.~W\c{a}s, 
Comput. Phys. Commun. {\bf 79}, 291 (1994).

\bibitem{ikaros-rho1}
M.~A.~Shifman and M.~B.~Voloshin, Sov. J. Nucl. Phys. {\bf 45}, 292 (1987);
V.~A.~Khoze, M.~A.~Shifman, N.~G.~Uraltsev, and M.~B.~Voloshin,  Sov.
J. Nucl. Phys. {\bf 46}, 112 (1987).

\bibitem{artuso}
M.~Artuso and E.~Barberio in \cite{pdg2002} [hep-ph/0205163].

\end{thebibliography}
\end{document}